\documentclass[twocolumn,showpacs,preprintnumbers,amsmath,amssymb]{revtex4}

\usepackage{graphicx}

\begin{document}

\title
{The static response and the Local-field factor of the 2-D electron fluid.
}

\author
{
 M.W.C. Dharma-wardana\cite{byline1}}
\author{Fran\c{c}ois Perrot$^{\ddag}$}
\affiliation{
Institute of Microstructural Sciences,
National Research Council of Canada, Ottawa,Canada. K1A 0R6\\
}
\date{\today}
\begin{abstract}
The static local-field factor (LFF) of the interacting 2-D  fluid is
calculated nonperturbatively using a mapping of the quantum
fluid to a classical Coulomb fluid 
$\lbrack$Phys. Rev. Let., {\bf 87}, 206 (2002)$\rbrack$.
The LFF for the paramagnetic fluid {\it differs markedly}
from expectations from
 standard  perturbation theory.
 Our
result for the LFF has a maximum close to 3$k_F$, while
perturbation methods yield a maximum near 2$k_F$.
The available quantum Monte Carlo data seem
to agree with our results.
These findings imply that  many effects, e.g., Friedel
oscillations, Kohn anomalies, effective electron-electron
interactions, etc., are affected by the strong correlations among 
anti-parallel spins in 2D.
%
%
\end{abstract}
%
\maketitle
%
{\it Introduction.}
The study of the uniform two-dimensional electron
 gas (2DEG) continues to bring out new and unexpected
properties. Its physics
 depends crucially on the  ``coupling parameter''
$\Gamma$ = (potential energy)/(kinetic energy)  which defines the
strength of the  Coulomb
interactions.
The $\Gamma$ for the 2DEG at $T=0$ and mean density $n$ turns out to be
equal to the mean-disk radius $r_s=(\pi n)^{-1/2}$ per electron.
The density parameter $r_s$, the
spin polarization $\zeta$ and the temperature
$T$ are the only physical variables that appear in this problem.
As the coupling constant increases, the 2DEG becomes more like
a ``liquid'', but here we continue to loosely
call it an electron ``gas''. An important property of the
2DEG which captures many of the exchange and correlation
effects and central to its physics is the
response function $\chi(k,\omega)$. It is  expressed
in terms of a zeroth-order response
 $\chi^0(k,\omega)$ and a local-field factor (LFF), also
 called a local-field correction (LFC),
denoted by $G(k,\omega)$.
\begin{equation}
\label{lffdef}
\chi(k,\omega)=\chi(k,\omega)^0/[1-V_k(1-G(k,\omega))]\chi(k,\omega)^0
\end{equation}
The LFF is also closely related to the vertex function $\Lambda(k,\omega)$.
The simplest static form, $G(k)$, is identical with $G(k,0)$
at $k=0$, and begins to differ from
$G(k,\omega)$ as $k$ or $\omega$ increases.
 In general, the
$G(k,\omega)$ for $\omega$ smaller than the plasma frequency
$\omega_p$ behaves like a static quantity. Hence
the use of a static form $G(k)$ is often adequate.
As such, considerable effort has been devoted to determining
 $G(k)$, using perturbation theory,  kinetic equation methods
\cite{rk,stls}, etc.
  A partially analytic, partially empirical, approach is often
implemented by invoking parametrized models constrained
to satisfy sum rules~\cite{iwamoto}. 
These sum rules 
invoke physical quantities which are
 usually obtainable only
from numerical simulations.
While these efforts have considerably extended the theory beyond the
random-phase approximation (RPA) response, its validity
is still restricted to the low-$r_s$ regime.

	We have shown in a number of recent papers~\cite{prl1,prb00,prl2}
that the static properties of the 2D and 3D electron systems can be calculated
{\it via} an equivalent {\it classical} Coulomb fluid having a
temperature $T_q$ such that it has the same correlation energy as the 
quantum system at the physical temperature $T=0$.
The ``quantum temperature'' $T_q$ was shown to be given by~\cite{prl2},
\begin{equation}
\label{2dmap}
t=T_q/E_F=2/[1+0.86413(r_s^{1/6}-1)^2]
\end{equation}
where $E_F=1/r_s^2$ is the Fermi energy in Hartree atomic units.
At finite temperatures, the classical-fluid temperature
$T_{cf}$ is taken to be $(T_q^2+T^2)^{1/2}$, as
discussed in Ref.~\cite{prb00}. The pair-distribution functions (PDFs)
of the classical fluid  are calculated using the 
hyper-netted-chain (HNC) equation \cite{hncref}
inclusive of bridge corrections.
A set of three coupled equations, for $g_{11}$, $g_{12}$ and $g_{22}$
for the two spin-species i=1,2 is solved.
This  {\it classical} mapping of
 quantum fluids within the HNC
was named  the CHNC.
 The static response function and the static
structure factor of a classical fluid are identical (except for numerical
 factors), and we also showed \cite{prl1,prb00} how this 
can be exploited for a very simple
 determination
of the static LFF of the  quantum electron fluid. In ref.~\cite{prb00} we 
 showed that various sum rules, e.g, the behaviour of G(k)
at large $k$ and its
relation to the ``on-top'' value $g(0)$ of the
 pair-distribution function (PDF) are satisfied by this ``classically''
determined $G(k)$, and that the results agreed well with,
e.g., the 3D-LFF of Utsumi and Ichimaru~\cite{utsumi}.

The application of the CHNC method to the 2DEG 
required the inclusion of short-range clustering effects 
(``bridge terms'') going  beyond the usual
 HNC approximation. These contributions seem to play a role 
 similar to the ``back-flow'' contributions used in 
 QMC simulations~\cite{kwon}.
 Such terms were necessary for $g_{12}$, while the
need was less stringent for $g_{11}$ since
the Pauli exclusion  blocked the clustering for identical spins.
However, it is well known that the compressibility sum rule is violated
by the HNC approximation, and that a bridge term is in principle
necessary~\cite{lado}, even for 
$g_{ii}$. While this shortcoming is not strongly felt in the calculated
distribution functions \cite{balutay}, and even less so in the energy, it needs to be
addressed if the small-$k$ limit of the LFF is to be correctly recovered.

	The objective of this work is to evaluate the LFF of the 2DEG
as a function of $r_s$ and show that it does {\it not} behave in the
manner expected from standard perturbation calculations~\cite{teter}
 using the
RPA screened Coulomb interaction. Such  calculations give
LFFs with a ``hump'' at 2$k_F$ due to the singular nature of $\chi(k=2k_F)$, 
 while we find that the interactions have moved
the hump towards  $\sim 3k_F$. A limited set of QMC data for the 2-D LFF is
available in the literature\cite{tosi,moroni} and seems to be in agreement with our findings.
If this is indeed the case,  effects like the
Kohn anomaly, Friedel oscillations, effective attractive
electron-electron  interactions\cite{teter} etc.,  would be {\it weaker} in 
2-D.
This emphasizes the need for further work in this field.

{\it The local-field factor.}
	The LFF was already defined in Eq.~\ref{lffdef}.
Here we are concerned with the static form $G(k)$, defined
 with respect to
a reference ``zeroth-order'' response function. It
is customary to use the Lindhard function $\chi(k)^0_L$ for this purpose.
However, a number of author~\cite{nicklasson}
 pointed out that another natural choice is to use the ``density-functional''
non-interacting form $\chi(k)^0$ containing the occupation numbers
 corresponding to the
{\it interacting} density. Richardson and Ashcroft(RA)~\cite{richash}
explicitly constructed a set of LFFs based on this $\chi(k)^0$, containing
the corrected occupation numbers. The calculation of RA was for the 3DEG, and
identified the simplest conserving 
set of diagrams in the perturbation expansion for the
RPA-screened interaction. This set (Fig. 2 of RA) is the same
 as selected earlier by
Geldart and Taylor~\cite{GT}
 who calculated the 
static 3D case. 
In all cases, dynamic or static,
the LFF
calculated from the diagrams is further restructured using quantum Monte-Carlo
data in several ways. In effect, most of the currently proposed LFFs, be they
for 2D (e.g., \cite{tosi})  or 3D (e.g, \cite{utsumi}),
 are in some sense only partially analytic  since
 they use the numerical simulation (QMC) 
 data and fit the behaviour at $k$=0, $k\to\infty$ and possibly 
 $S(k)$ etc., to form the  LFF.
 This truly  emphasizes
 the 
 difficulty and delicateness involved in the determination of the LFF.

	The CHNC also uses the  QMC correlation-energy $E_c(r_s)$,
at the initial stage
of constructing the ``quantum temperature'' $T_q(r_s)$ of the classical fluid
at zero temperature ($T=0$). 
From then on we have a self-contained method for obtaining all the  static
properties including $E_c(r_s,\zeta,T)$ and its gradients, as well as
$g(r)$, $S(k)$ etc.
The CHNC also provides a {\it very simple} formula for the LFF, via the
classical-fluid LFF. Unlike in the quantum case,
for a classical fluid, $\chi(k)$ is
directly related to the structure factor.
\begin{equation}
S_{ij}(k)=-(1/\beta)\chi_{ij}(k)/(n_in_j)^{1/2}
\label{StoChi}
\end{equation}
Hence, taking the paramagnetic case for simplicity, 
\begin{equation}
\label{lfcCS}
V_{c}(k)G(k)=V_{c}(k) -\frac{T_{cf}}{n}
\Bigl\lbrack\frac{1}{S(k)}-\frac{1}{S^0(k)}\Bigr\rbrack
\end{equation}
Here $T_{cf}$ equals $T_q$
if the physical temperature $T$ = 0.
In CHNC, and hence in these expressions,
 the $\chi^0(k)$  and $S^0(k)$ are based on a
Slater determinant, while the
 Lindhard function is applicable to the non-interacting case without
antisymmetrization of the wavefunction.

The above expressions show that the LFF is immediately available if the
interacting and noninteracting structure factors are known. These are
explicitly known since the HNC+Bridge equations for the classical
fluid yield the PDFs $g_{ij}(r)$ and  Fourier transforms of which
directly yield the $S_{ij}(k)$ needed here. Alternatively, any other
source of $S(k)$, e.g., QMC, may be used, while $S^0(k)$ for the
2DEG is analytically known.
Equation~\ref{lfcCS} involves a difference between the
inverses of $S(k)$ and $S^0(k)$, and hence it is very desirable to calculate 
them to the same accuracy, since their
difference has to cancel the Coulomb potential which becomes large
 as $k\to 0$.  
The Coulomb potential $V_c(r)$
for point-charge electrons is  $1/r$.
However, the classical electron at
the temperature $T_{cf}$ is
localized to within a thermal wavelength. Thus,
we use the ``diffraction
corrected'' form~\cite{prl2}
\begin{eqnarray}
\label{potd}
V_{c}(r)&=&(1/r)[1-e^{-rk_{th}}]\\
V_{c}(k)&=&2\pi[k^{-1}-(k_{th}^2+k^2)^{-1/2}]
\end{eqnarray}
The $k_{th}$ was  determined by numerically
 solving the Schrodinger equation
 for a pair
of 2-D electrons in the potential $1/r$ and calculating the electron density
in each normalized state~\cite{pdwkth}.
It was found that
 $$k_{th}/k^0_{th}=1.158T_{cf}^{0.103}$$
 where $T_{cf}$ is in a.u., Here  $k^0_{th}$
is the de Broglie thermal wavevector $(2\pi m^*T_{cf})^{1/2}$,
with $m^*$, the reduced mass of the scattering electron-pair,
equal to 1/2.

The introduction of the de Broglie wavevector $k_{th}$ adds a new
$k$-scale which competes with the Fermi wavevector
$k_F$  of the non-interacting problem.
However, $k_{th}$ becomes equal to about $k_F$ only when we
approach the Wigner crystallization ( $r_s \sim 35$ ) regime.
Hence, as far as this study is concerned, the critical
wavevector
$k_{c}$  which separates the small-$k$ region from the large-$k$ 
region will be taken to be 2$k_F$.
Although Eq.~\ref{lfcCS} is already very simple to compute, the 
 explicit cancellation of $V_{c}$ by the terms in the $1/S-1/S_0$
can be realized as follows.
For the paramagnetic case, we have:
\begin{eqnarray}
S(k)&=&1+nx[h_{11}(k)+h_{12}(k)]\\
    &=&1/[1-nx\{c_{11}(k)+c_{12}(k)\}]
\end{eqnarray}
Here we have used the Orstein-Zernike relations and the
direct ($c_{ij}$) and total ($h_{ij}$) correlation functions,  while
 $x$=1/2 is the fractional composition.
 The pair-distribution
 function~\cite{prl2} in CHNC has the form
\begin{eqnarray}
g_{ij}(r)&=&e^{[-\beta\phi_{ij}(r)-c_{ij}(r)+h_{ij}(r)+B_{ij}(r)]}\\
\phi_{ij}(r)&=&P_{ij}(r)\delta_{ij}+V_{c}(r)
\end{eqnarray}
Here $P_{ij}$ is an effective potential for the Pauli-exclusion effect,
and is such that, when used in the HNC
equations, reproduces the non-interacting PDF, $g^0(r)$.
On defining  modified (short-ranged) direct-correlation functions:
\begin{equation}
\hat{c}_{ij}(r)=c_{ij}(r)+\beta \phi_{ij}(r)
\end{equation}
we can express the LFF in a form where 
$V_{c}(k)$ has been explicitly cancelled out.
\begin{equation}
\label{lfcFP}
G(k)=Tx[\hat{c}_{11}+\hat{c}_{12}+ 
B_{11}(k)+ B_{12}(k)-\hat{c}^0_{11}]/V_{c}(k).
\end{equation}
This is free of numerical inaccuracies at small-k, and contains the
bridge terms that may be used to recover the compressibility
sum rule etc. On the other hand, this requires explicit
forms for the bridge functions (which can be constructed using the
hard-disk model bridge function~\cite{prl2}).
However, here we propose a much simpler, {\it practical}, approach
based on using Eq.~\ref{lfcCS} which requires only
the  $S(k)$ and $S^0(k)$ which may be available from
other sources beside the CHNC.

The 2-D compressibility sum rule \cite{iwamoto} leads to the following
small-$k$ form for the LFF.
\begin{eqnarray}
\label{lffq0}
G(k)&=&C_0\,k\;\;k\to 0\\
C_0&=&\frac{1}{\pi}+\frac{1}
{4}\alpha r_s^2\Bigl\lbrack \frac{\partial E_c}{\partial r_s}
-r_s \frac{\partial^2 E_c}{\partial r_s^2}\Bigr\rbrack
\end{eqnarray}
Here $E_c$ is the correlation energy per particle in atomic units,
and $\alpha=1/\surd{2}$. Several
 fit formulae (based on QMC data,
\cite{TC,rapi,atta}) are available for $E_c(r_s)$.
Hence $C_0$ can be calculated easily.
 Equation~\ref{lffq0}
is valid only for small-$k$ where $G(k)$ is linear
in $k$. Based on more extensive calculations using Eq.~\ref{lfcFP}, 
we assume a linear interpolation for the small-$k$ region $k_1$ to $k_c$,
where $k_1$ is the smallest value of $k$ in our $S(k)$
tabulation. That is, we replace the form:
\begin{equation}
\label{allk}
G(k)=1-\frac{T_{cf}}{n V_{c}(k)}
\Bigl\lbrack\frac{1}{S(k)}-\frac{1}{S^0(k)}\Bigr\rbrack; \;\; all\;k
\end{equation}
by the constrained form:
\begin{subequations}
\begin{eqnarray}
\label{linearized}
G(k_1)&=&C_0\,k_1 \nonumber \\
G(k)&=&\Bigl\lbrack G(k_c)-G(k_1) \Bigr\rbrack
\frac{k-k_1}{k_c-k_1}\;\; k < k_c. \label{subeq:1} \\
G(k)&=&1-\frac{T_{cf}}{n V_{c}(k)}
\Bigl\lbrack\frac{1}{S(k)}-\frac{1}{S^0(k)}\Bigr\rbrack
\;\;k>k_c.\label{subeq:2}
\end{eqnarray}
\end{subequations}

Here we choose $k_c=2k_F$. In practice, $k_c$ is chosen to be the
k-point closest to 2$k_F$ in the $S(k)$ tabulation which gives a continuous
accord between the low-$k$ and high-$k$ regimes, and hence $k_c/k_F$ may be,
e.g., 1.9 or  2.1.
%
\begin{figure*}
\includegraphics*[width=14.0cm, height=14.0cm]{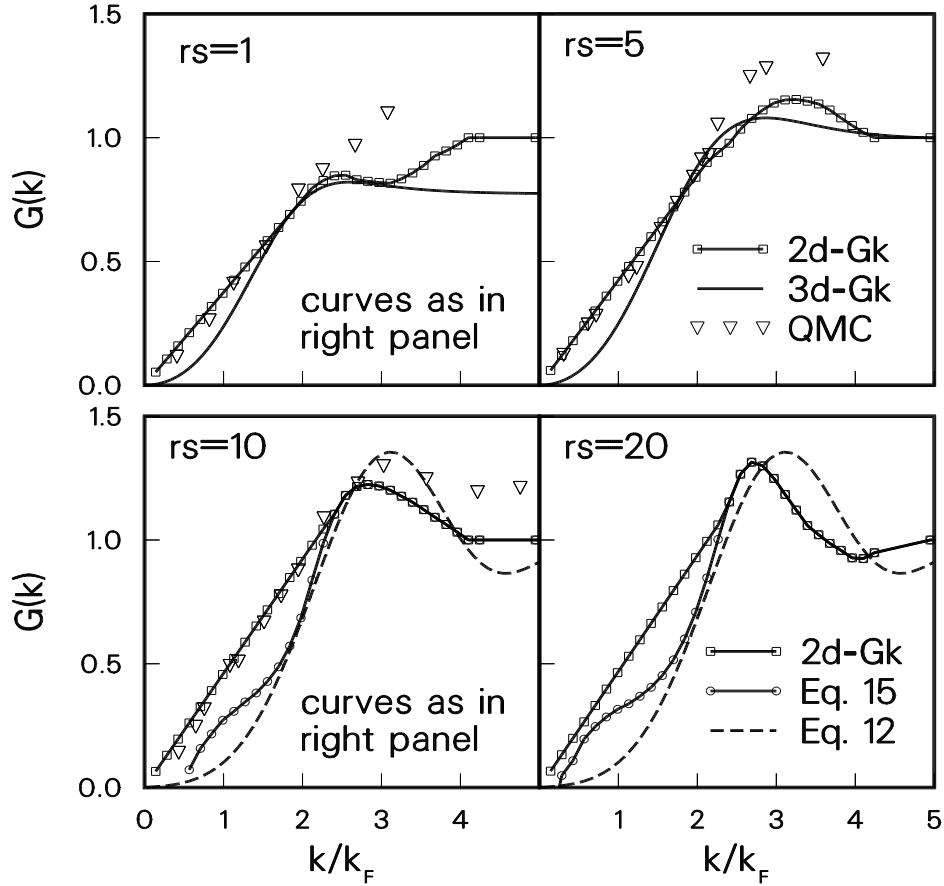}
\caption
{The local-field factors  calculated  with linearization (squares),
 Eq.~\ref{linearized},
and without linearization (circles), Eq.~\ref{allk},
are shown for $r_s$= 10 and 20.
For $r_s$=1, 5  and 10 QMC data (triangles) extracted from Fig. 1 of
ref~\cite{tosi} are shown, and have a $q$-dependent asymptotic form,while CHNC tents to $1-g(0)$ for large $q$.
 The solid continuous line in the
panels for $r_s$=1 and 5 are the 3-D LFFs calculated from CHNC.
}
\label{lfcfig1}
\end{figure*}

%
{\it Results.}
  We now consider the LFF for $r_s=1,5,10$ and 20 calculated using
Eq.~\ref{lfcCS}, and displayed in Fig.~\ref{lfcfig1}. 
We have also displayed the 3DEG local-field factor
 for $r_s=1$ and 5. The 3D-LFF shows a weak maximum 
 near 2$k_F$ and tends to $1-g(0)$ for large $k$.
The QMC data (triangles) are extracted from  Fig.~1 of ref.\cite{tosi},
and are not available for $r_s$=20.
Note that these do not tend to the $1-g(0)$ but has a linear-$q$
dependence for asymptotically  large -$q$,
  consistent with a zeroth-order Lindhard response.  
For $r_s=1$, the thermal
wavevector ($k_{th}/k_F =2.88$) is larger than $2k_F$. 
Hence the region where
$G(k)$ is linear in
$k$ may extend well beyond 2$k_F$. However, we have retained a linearization
up to $2k_F$ which joins with the $k\,>\,2k_F$ behaviour given from CHNC. 
Also, the  extended linearity in the QMC is partly due to its
large-$q$ asymptotic behaviour.
For  $r_s$=5 strong-coupling effects show up and 
  a broad hump-like structure 
near $k/k_F \sim 3$ is manifest. When we go to $r_s=10$ this
 behaviour is more pronounced, and sharpens even more
  for $r_s$=20.
The required large-$k$ behaviour, $G(k)\to 1-g(0)$, of the LFF is
automatically satisfied by CHNC and hence we do not impose it.
 The difference between $1/S(k)$ and $1/S^0(k)$  required in Eq.~\ref{allk}
 increases for strong
 coupling, and  the direct method of calculating the LFF, using Eq.~\ref{lfcCS},
 instead of Eq.~\ref{lfcFP} becomes adequate. Thus we compare
 the LFF calculated without imposing the compressibility
 sum rule, and using $S(k)$ from two different sources and
 different (but equivalent) formulations. For the cases $r_s=10$ and 20, we
 have given (dashed line)  LFFs evaluated purely using the
  $S(k)$ and $S^0(k)$ derived
 from CHNC, and using  Eq.~\ref{lfcFP}. These include only the $B_{12}$ bridge
 term. The linear form based on the compressibility sum rule {\it has not been}
 imposed. We also present the LFFs (solid lines with circles) obtained 
 using the $S(k)$ given in 1989 by
 Tanatar and Ceperely\cite{TC}, together with the linearized forms 
 (solid lines with squares)
 resulting from them. These results show that the maximum around 3$k_F$
 is obtained irrespective of the method used, while the small-$k$
 region is clearly sensitive to the numerical procedures used.
 More details about the CHNC method and on-line access to our codes may
 be obtained at our website\cite{web}.
 
 {\it conclusion}
 Unusual features of the 2D-LFF not found in the 3D-case, and unexpected from
 perturbation theory, have been revealed via the CHNC method. These results
 seem to be in agreement with the limited QMC results currently
available. They should
 stimulate further work on the important problem of the 2D response
 using non-perturbative
 methods.

\end{document}